\documentstyle[amssymb,prb,aps,preprint,floats]{revtex}

\input epsf

\begin{document}
\title{{\sc On the Problem of Predicting Inflationary Perturbations }}
\author{Limin Wang$^a$, V.F. Mukhanov$^b$ and Paul J. Steinhardt$^a$}
\address{$^a$Department of Physics and Astronomy\\
University of Pennsylvania \\
Philadelphia, Pennsylvania 19104 USA}
\address{$^b$Institut f\"{u}r Theoretische Physik, ETH Zurich, CH-9093 Zurich,\\
Switzerland}
\maketitle

\begin{abstract}
We examine the theoretical foundations of standard methods for computing
density perturbations in inflationary models. We find that: (1) the
time-delay formalism (introduced by Guth and Pi, 1982) is only valid when
inflation is well-described by the de Sitter solution
 and the equation-of-state is nearly unchanging;
and, (2) the horizon-crossing/Bessel approximation extends to
non-exponential inflation, but only if the equation-of-state is changing
slowly. Integration of the gauge-invariant perturbation equations
mode-by-mode is the only method reliable for general models. For models with
rapidly varying equation-of-state, the correction leads to significantly
different predictions for the microwave background anisotropy. An important
corollary is that methods proposed for ``reconstruction" of the inflaton
potential from anisotropy data are unreliable for general models.

\end{abstract}

\baselineskip 25pt

\newpage

One of the most important predictions of inflationary cosmology\cite
{Guth,Linde,AS} is that quantum fluctuations of the inflaton field grow into
cosmological energy-density perturbations.\cite{MC,BST,GP,Hawking,Starobinskii}.
In this paper, we analyze and compare the standard methods for computing
perturbation spectra in inflation. 
This consideration is motivated by the need to have 
theoretical predictions which match the precision anticipated in forthcoming
measurements.
We find that the simplest, most commonly used methods are  approximations 
with narrow ranges of validity.  The only reliable method for general 
potentials is the gauge invariant method\cite{BST,M1,M2} in which 
the equation-of-motion for the perturbation must be integrated for each    
mode.  

 The paper has several disparate components which we have organized by 
section for the convenience of the reader: (1) a review of the 
``exact"  gauge invariant methods\cite{BST,M1,M2} 
with attention to some subtleties which have  caused
confusion in past literature; (2) an analysis of the time-delay
formalism   demonstrating that 
it is  valid only when inflation is nearly de Sitter
(exponential inflation); (3) an analysis of the horizon-crossing/Bessel
function approach,\cite{BST,Starobinskii,Hawking,SL} 
which show that it extends to non-exponential
inflation but only if the equation-of-state is changing slowly; 
(4)  example (Figure 1)  of how the 
approximate methods can lead to large
errors in computation of cosmic microwave background (CMB) anisotropy in cases of
 rapidly varying equation-of-state;
 (5) a summary in terms of conditions on the 
   inflaton potential for applying the time-delay and horizon-crossing
approximations
and discussion of implications for
 ``reconstructing" the inflaton potential and
constraining cosmological parameters from  (CMB)
anisotropy and large-scale structure data.

\noindent {\bf Exact Method:}\cite{BST,M1,M2} In this paper, we consider
the case of a single inflaton field; multi-field inflation introduces other
subtleties that  we will discuss in a future paper.\cite{Fut2}
The most general form for the
metric with linear scalar perturbations is\cite{Bardeen,MFB}: 
\begin{equation}
ds^{2}=a^{2}(\tau )\{(1+2\phi )d\tau ^{2}-2B_{|i}dx^{i}d\tau -[(1-2\psi
)\delta _{ij}+2E_{|ij}]dx^{i}dx^{j}\}  \label{metric}
\end{equation}
where $\phi ,B,\psi $ and $E$ are arbitrary functions of space and time. A
gauge-invariant combination of these variables is the gravitational
potential, 
\begin{equation}
\Phi =\phi +\frac{1}{a}[(B-E^{\prime })a]^{\prime },
\end{equation}
where prime means derivative with respect to conformal time $\tau $. The
potential $\Phi $ is simply related to the anisotropy of the CMB on large
angular scales via 
\begin{equation}
\frac{\delta T}{T}\simeq \frac{1}{3}\Phi .
\end{equation}
However, $\Phi $ is not the most convenient variable for tracing the
generation of perturbations by quantum fluctuations. For this purpose, it is
useful to introduce a second gauge invariant quantity\cite{M2}
\begin{equation}  \label{vvv}
v \equiv a\left( \delta \varphi +\frac{z}{a}\psi \right)
\end{equation}
where $\delta \varphi $ is the perturbation in the scalar inflaton field: $
\varphi _{total}(\vec{r},t)=\varphi_0 (t)+\delta \varphi (\vec{r},t)$.  During
inflation, the variable $z$ is 
\begin{equation}  \label{zzz}
z \equiv a \sqrt{\epsilon }
\end{equation}
where $\epsilon $ is the variable that characterizes the equation-of-state: 
\begin{equation}
\epsilon =\frac{3}{2}\left( \frac{\rho +p}{\rho }\right);
\end{equation}
for an inflaton with potential $V(\varphi)$,  pressure $p=\frac{1}{2}\dot{
\varphi}^{2}-V(\varphi )$, and energy density $\rho =\frac{1}{2}\dot{\varphi}
^{2}+V(\varphi )$. Hence, $z=a\dot{\varphi}/H$ , where $H=\dot{a}
/a=a^{\prime }/a^{2}$ is the Hubble constant and dot denotes the derivative
with respect to the physical time $t=\int ad\tau .$ We use everywhere the
units where $4\pi G=1.$ In the post-inflationary phase when the universe is
filled with hydrodynamical matter, the definition of $z$ in Eq.~(\ref{zzz})
is replaced by $z= (a/c_s) \sqrt{\epsilon}$ where $c_s$ is the speed of
sound and, in the definition for $v$ in Eq.~(\ref{vvv}), $\delta \varphi$ is
replaced by the potential of the peculiar velocities in the matter.

By expanding $v(x,\tau )$ in Fourier modes with comoving wavenumbers $k$,
the equation -of-motion for Fourier component $v_{k}$ becomes (after lengthy
computation): 
\begin{equation}
v_{k}^{\prime \prime }+\left( k^{2}-\frac{z^{\prime \prime }}{z}
\right) v_{k}=0  \label{dynamics}
\end{equation}
($k^2$ is replaced by $c_s^2 k^2$ in the post-inflationary stage).
The canonical variable $v$ used in deriving quantum fluctuations beginning
from an action describing the scalar field coupled to 
Einstein gravity;\cite{M2} and,
it can also be simply related to the gravitational potential $\Phi $
\thinspace via the constraint equation derived from the $0-0$ component of
the Einstein equations: 
\begin{equation}
\Phi _{k}=-\frac{H}{k^{2}a}z^{2}\left( \frac{v_{k}}{z}\right) ^{\prime }
\label{phi}
\end{equation}

The rigorous method to compute the perturbation spectrum is to solve the
second order equation for each $v_{k}$ using Eq.~\ref{dynamics}, beginning
from when the given wavelength is small compared to the horizon to when it
grows much larger than the horizon. To characterize the perturbations we use
the power spectrum of $\zeta =v/z$ defined as: 
\begin{equation}
P_{\zeta }(k)=\frac{k^{3}}{2\pi ^{2}}\left| \frac{v_{k}}{z}\right| ^{2},
\label{eq:def}
\end{equation}
This power spectrum can be easily related with the power spectrum of the
gravitational potential if we use Eq.~(\ref{phi}), assuming $z$ does not
vanish. Note that all of the equations above describe the perturbations not
only during inflation, but also in the post-inflationary (hydrodynamical)
Universe. After inflation, $\epsilon ={\cal O}(1) $ and for the long
wavelengths perturbations ($c_{s}^{2}k^{2}<<\left| z^{\prime \prime
}/z\right| $), $\Phi _{k}\sim {\cal O}(1)\zeta _{k}$. Hence, $P_{\zeta}$ is,
up to a constant of order unity, the power spectrum of the gravitational
potential that is of interest for computing perturbations of the CMB and the
formation of large-scale structure. In this paper, our focus is on the
fluctuations in $\zeta$ and their sensitivity to the equation-of-state, $
\epsilon$, during the  inflationary epoch.

We note that some discussions of the exact approach improperly characterize $
\zeta $ as a ``conserved quantity.'' Indeed, an approximate conservation law
can be derived for $\zeta $ using Eq.~(\ref{phi}) and taking the
long-wavelength limit $k\rightarrow 0$. However the ``conservation law'' is
not strictly true for finite $k$ and neglecting that fact can lead to some
confusion.\cite{Grishchuk} For example, integrating the equation, 
\begin{equation}
\frac{v}{z}\equiv \zeta =\frac{\Phi ^{\prime }/{\cal H}+\Phi }{\epsilon }
+\Phi ,  \label{zeta}
\end{equation}
which follows from $0-i$ Einstein equations, one can obtain $\Phi $ for a
given $\zeta $; in the long-wavelength limit ({\it i.e.,} to lowest order in 
$k$), one appears to obtain non-physical, extra constants of integration. To
see that there are not additional integration constants, one should use
constraint equation Eq.~(\ref{phi}) and keep terms to order $k^{2}$. The
``extra'' constants are then fixed in terms of the ``old'' ones.

\noindent {\bf Time-delay Formalism}: 
The time-delay formalism\cite{GP} is one of the methods originally
introduced   to compute
the energy density perturbation spectrum  at the end of inflation.
In this method, the perturbations are related to the proper time 
delay $\delta \tau(x)$ between when inflation ends at position $x$
compared to the spatial average.  Here, we wish to show that this 
method is limited to cases in which the inflaton potential can 
be treated as nearly flat. 
To be sure, 
the time-delay formalism is a more intuitive  derivation
than the gauge-invariant approach
 and treating the potential as flat suffices
as a crude estimate for some models.  However, if one is interested
in a  rigorous treatment or 
 more general models, including typical models of chaotic, natural,
and power-law inflation, 
 the time-delay formalism fails.

The time-delay  method was originally presented for a 
toy model in which the inflationary phase is  well-described by
the  de Sitter  solution
terminated by an instantaneous  
transition to a hydrodynamical stage. 
The  aspects of the toy model which are essential to the time-delay
approach are:
 (a) 
the metric perturbations can be completely ignored during the strictly
de Sitter
stage,  and (b) the time delay for the perturbations with comoving wave number $k$,
defined as $\delta \tau _{k}=\delta \varphi _{k}(t)/\dot{\varphi}_{0}(t)$,
approaches  a time-independent constant when the perturbation stretches well
beyond the  horizon. 

Formally, the time-delay formalism cannot be made rigorous for any model,
as it follow from the equation for the background $\dot{H}=-\dot{\varphi}
_{0}^{2}$.
The strict de Sitter limit, $\dot{H}=0$, requires that
$\dot{\varphi_0} =0$, in which case $\delta \tau _{k}=\delta \varphi _{k}(t)/\dot{\varphi}_{0}(t)$ is divergent.  However, there is a limited range of 
models for which the formalism gives the leading order contribution.
This range is what we wish to clarify here.   If one assumes the 
slow-roll approximation in which $\ddot{\varphi}$ can be ignored in the
equation-of-motion for $\varphi$, then 
\begin{equation} \label{back}
\dot{\varphi}_0 \approx - V,_{\varphi}/3 H,
\end{equation}
where $V,_{\varphi} = dV/d\varphi$.
The perturbation $\delta \phi$, satisfies the perturbed, linearized 
Klein-Gordon equation:
\begin{equation}
\delta \ddot{\varphi}+3H\delta \dot{\varphi}-\frac{1}{a^{2}}\nabla
^{2}\left( \delta \varphi +a\dot{\varphi}_{0}(B-a\dot{E})\right)
+V,_{\varphi \varphi }\delta \varphi +2V,_{\varphi }\phi -\dot{\varphi}_{0}(
\dot{\phi}+3\dot{\psi})=0, \;  \label{scfper}
\end{equation}
which is  supplemented by the $0-i$ Einstein
equation
\begin{equation}
\dot{\psi}+H\phi =\dot{\varphi}_{0}\delta \varphi.   \label{Eeq}
\end{equation}
In the long-wavelength and slow-roll limits,  $\ddot{\varphi}$ and
time- and spatial-derivatives of the metric parameters can be dropped:
\begin{equation}
3 H \delta \dot{\varphi} + \left(V,_{\varphi \varphi } - \frac{(V,_{\varphi })^2}{V} \right) \delta \varphi =0.
\end{equation}
(In dropping these terms, we ignore the decaying modes 
and assume a generic gauge in which none of the
variables characterizing the perturbations are suppressed compared to others
by the gauge choice.)
The  solution for $\delta{\varphi}$ 
\begin{equation} \label{eqvar}
\delta \phi = C \frac{V,_{\varphi}}{V},
\end{equation}
where the constant $C \equiv (H V,_{\varphi}/V)_0$  can be expressed in terms
of $V$ and $H$ evaluated at horizon-crossing for the given mode; 
this choice of $C$ guarantees
that $\delta \varphi \rightarrow H$ at horizon-crossing
as expected in the de Sitter limit.  

The key point is that, from Eqs.~(\ref{back}) and (\ref{eqvar}), we
find that 
\begin{equation}
\delta \tau  = \frac{\delta \varphi}{\dot{\varphi_0}}  = \frac 13 C \frac{H}{V} \propto \frac{1}{\sqrt{V}}.
\end{equation}
Unless $V$ is independent of $\varphi_0(t)$, $\delta \tau$ depends on 
time which is inconsistent with an essential criterion of 
the time-delay formalism.
Since $V$ is always $\varphi$-dependent in practice, the time-delay formalism
is, at best, a lowest order approximation.  Even so,
it should only be applied if the $V$ is extremely flat  and $H$ is
nearly constant over the range of 
e-folds of physical interest, typically the last 60 e-folds of inflation.
If $H$ is nearly constant over the 60-folds, then 
the integral of    $\epsilon =d (1/H)/dt$ over the last 60 e-folds
must be much less than $1/H$;  if the integral is to be less by a 
factor  $\delta \ll 1$, then we require
\begin{equation}
\begin{array}{rcl}  \label{time}
\epsilon & \le & \frac{\delta}{60}, \\
\frac{d \, {\rm ln}\, \epsilon}{dN} & \le & \frac{\delta}{60}, \\
\epsilon \frac{d \, {\rm ln}\, \epsilon}{dN} & \le & \frac{\delta}{3600};
\, \, \ldots
\end{array}
\end{equation}
where  $\epsilon$ must satisfy these constraints during the last 60 e-folds,
$N$ is the number of e-folds, and $\ldots$ refers to
analogous constraints on higher order derivatives.
This represents a narrow set of  models which excludes common power-law and
chaotic inflationary models.

\noindent {\bf Horizon-crossing/Bessel Approximation}: The
horizon-crossing/Bessel approximation is based on the exact 
gauge invariant method but circumvents mode-by-mode
integration. A recent review\cite{Lidsey} discusses prior work and 
contains references.  The perturbation
amplitude for a given mode as it enters the horizon in the post-inflationary
epoch is expressed in terms of the amplitude when the mode crosses beyond
the horizon during inflation. To obtain the latter amplitude, the solution
of Eq.~(\ref{dynamics}) for the non-decaying mode in the long-wavelength
limit, $k^{2}\ll |z^{\prime \prime }/z|$, 
\begin{equation}
v_{k}\longrightarrow C(k)z  \label{eq:outside}
\end{equation}
is matched to the solution in the short wavelength limit, $k^{2}\gg
|z^{\prime \prime }/z|$, 
\begin{equation}
v_{k}\longrightarrow \frac{1}{\sqrt{2k}}e^{-ik\tau }.  \label{eq:inside}
\end{equation}
at horizon-crossing, $k={\cal O}(1)aH$. (The normalization in Eq. (\ref
{eq:inside}) follows from the fact that $v$ is the quantum canonical
variable and $c_{s}=1.$) The matching condition determines $C(k)$; namely, $
|C(k)|={\cal O}(1)/\sqrt{2k}z$ where $z\equiv a\dot{\varphi}/H$ evaluated
when $k={\cal O}(1)aH$; hence, the power spectrum is 
\begin{equation}
P_{\zeta }(k)\rightarrow \frac{k^{3}}{2\pi ^{2}}|C(k)|^{2} =
\left( \frac{H^{4}}{\dot{\varphi}^{2}}\right) _{0} \times
{\cal O} (1)  \label{ex1}
\end{equation}
in the long-wavelength limit, where the subscript $0$ means that $H$ and $
\dot{\varphi}$ are evaluated at $k=aH$ precisely. The ${\cal O}(1)$
ambiguity reflects the fact that the actual matching condition is not so
precise.

The Bessel approximation is an improvement of the horizon-crossing
approximation intended to resolve the ambiguity by replacing simple matching
at $k \sim aH$ with a Bessel function approximant to Eq.~(\ref{dynamics})
valid in a range of wavelengths around $k^2=|z^{\prime\prime}/z|$ and then
matching this solution to the long- and short-wavelength limits. More
accurately capturing the behavior of the exact solution near $
k^2=|z^{\prime\prime}/z|$ is important to our purpose because it is the
integration of Eq.~(\ref{dynamics}) over this wavelength regime that is
sensitive to the time-variation of the equation-of-state, $\epsilon(t)$. 
Hence, the Bessel approximation not only replaces the ${\cal O}(1)$ in Eq.~(
\ref{ex1}) with a known function, but specifically a function of $\epsilon$.

The horizon-crossing/Bessel
approximation appears at first glance  to be a leading order expression in
a systematic expansion that can be extended to arbitrarily high accuracy.
Here we show that, in actuality, it is obtained by matching long-,
intermediate- and short-wavelength solutions. In particular, the
horizon-crossing/Bessel approximation assumes that the amplitude of a mode
is determined by a conditions within a small range of e-folds around
horizon-crossing for the given mode. If the equation-of-state is changing
rapidly, this assumption breaks down.

The ratio $z^{\prime\prime}/z$ in Eq.~(\ref{dynamics}) can be re-expressed
in terms of $\epsilon$: 
\begin{equation}  \label{first}
\frac{z^{\prime \prime }}z=2H^2a^2\left( 1-\frac \epsilon 2-\frac 34\frac{
d\ln \epsilon }{dN}+\frac 14\epsilon \frac{d\ln \epsilon }{dN}+\frac 18
\left( \frac{d\ln \epsilon }{dN}\right) ^2+\frac 14\frac{d^2\ln \epsilon }{
dN^2}\right),
\end{equation}
where $0< \epsilon(N) < 1$ during inflation. Here the conformal time
variable $\tau$ has been replaced with $N$, the remaining number of e-folds
until the end of inflation: the differential $dN$ satisfies $dN=- a H d\tau $
. In Eq.~(\ref{dynamics}), the cross-over between short-wavelength and
long-wavelength behavior occurs at $k^2 \sim| z^{\prime\prime}/z|$. Note
that this corresponds to horizon-crossing, $k\sim aH$, only provided that $
\epsilon(N)$ does not change too rapidly, {\it e.g.,} ($d \, {\rm ln}\,
\epsilon/dN$, $d^2 \, {\rm ln}\, \epsilon/dN^2$)~$\le {\cal O}(1)$). This is
a necessary but insufficient condition for the horizon-crossing
approximation to be valid.

Assuming this condition is satisfied, the solution to Eq.~(\ref{dynamics})
for a given mode $k$ can be expressed as a Bessel function about $k=aH$. To
see this, it is useful to replace $N$ with 
\begin{equation}
x\equiv \ln (Ha/k)=\ln (\lambda _{ph}/H^{-1})  \label{def1}
\end{equation}
where $\lambda _{ph}=a/k$ is the physical wavelength of the perturbation
with comoving scale $k.$ The variable $x$ is roughly the number of e-folds
after the given mode crosses the horizon during inflation; it is equal to
zero at the moment of horizon-crossing, positive after horizon-crossing, and
negative before horizon-crossing. Then, if we replace $v$ with 
\begin{equation}
\stackrel{\sim }{v}=(1-\epsilon )^{1/2}(\exp (x/2))v  \label{def2}
\end{equation}
Eq.~(\ref{dynamics}) takes the form 
\begin{equation}
\begin{tabular}{l}
$\frac{d^{2}\stackrel{\sim }{v}}{dx^{2}}+\left[ \frac{\exp (-2x)}{
(1-\epsilon )^{2}}-\frac{1}{4}\left( \frac{3-\epsilon }{1-\epsilon }\right)
^{2}-\frac{3}{2}\frac{d\ln \varepsilon }{dx}+\frac{1}{2}\frac{d\ln
(1-\epsilon )}{dx}-\frac{1}{4}\left( \frac{d\ln \epsilon (1-\epsilon )}{dx}
\right) ^{2}-\frac{1}{2}\frac{d^{2}\ln \epsilon (1-\epsilon )}{dx^{2}}
\right] \stackrel{\sim }{v}=0$
\end{tabular}
\label{bess}
\end{equation}

If $\epsilon=$~constant then this equation is reduced to a Bessel equation
with exact solutions given by $\widetilde{v} \sim H^{(1,2)}_\nu \left( \frac{
e^{-x}}{1-\epsilon }\right)$ where 
\begin{equation}
\nu =\pm \frac 12\left( \frac{3-\epsilon }{1-\epsilon }\right)
\end{equation}
However, $\epsilon=$~constant is not realistic for cosmology since the
equation-of-state must change near the end of inflation in order to return
to the ordinary, Friedmann-Robertson-Walker expansion rate. The standard
approach has been to solve Eq.~(\ref{bess}) approximately near the horizon
crossing point $x=0$ where 
\begin{equation}
\epsilon (x)=\epsilon _0+\epsilon _0\left( \frac{d\ln \epsilon }{dx}\right)
_0x+...  \label{expan}
\end{equation}
The subscript 0 is used to symbolize horizon-crossing for the given mode,
which is equivalent to evaluating at $x=0.$ The solution will be an
approximation in which the small parameters are $\epsilon _0$ and $x$
-derivatives, $(d.../dx)_0$. More precisely, $\ln \, \epsilon _0$ is treated
as a zeroth-order quantity; $\epsilon _0$ and $(d\ln \epsilon /dx)_0-$ are
first order quantities; $\epsilon _0^2,$ $\epsilon _0 \, (d\ln \epsilon
/dx)_0,(d\ln \epsilon /dx)_0^2$ and $(d^2\ln \epsilon /dx^2)_0$ are second
order quantities; etc. Note that $d\ln (1-\epsilon )/dx=- (\epsilon/
(1-\epsilon)) (d\ln \epsilon /dx)$ is second order.

Substituting (\ref{expan}) in Eq.~(\ref{bess}), we obtain: 
\begin{equation}
\begin{tabular}{l}
$\frac{d^{2}\stackrel{\sim }{v}}{dx^{2}}+\left[ \frac{\exp (-2x)}{
(1-\epsilon _{0})^{2}}-\frac{1}{4}\left( \frac{3-\epsilon _{0}}{1-\epsilon
_{0}}\right) ^{2}-\frac{3}{2}\left( \frac{d\ln \epsilon }{dx}\right)
_{0}+R_{2}(x)\right] \stackrel{\sim }{v}=0$
\end{tabular}
\label{apbesse}
\end{equation}
where $R_{2}(x)$ denotes the second and higher order terms, 
\begin{equation}
\begin{array}{rcl}
R_{2} &=&\left( \frac{1}{2}-\frac{2e^{-2x}}{\left( 1-\epsilon _{0}\right)
^{2}}x+\frac{3-\epsilon _{0}}{\left( 1-\epsilon _{0}\right) ^{2}}x\right)
\left( \frac{d\ln (1-\epsilon )}{dx}\right) _{0}  \label{corr} \\
&&-\frac{1}{4}\left( \frac{d\ln \epsilon }{dx}\right) _{0}^{2}-\frac{1}{2}
\left( 1+3x\right) \left( \frac{d^{2}\ln \epsilon }{dx^{2}}\right)
_{0}+O(\epsilon _{0}^{3},...)  \nonumber
\end{array}
\end{equation}
The initial conditions are fixed by the assumption that the short-wavelength
behavior should be as in a flat Minkowski vacuum with only positive
frequencies, resulting in the solution 
\begin{equation}
\widetilde{v}=C_{k}H_{\nu }^{(1)}(\xi )  \label{sol}
\end{equation}
where $\xi \equiv \exp (-x)/(1-\epsilon _{0}),$ and 
\begin{equation}
\nu =\frac{1}{2}\left( \left( \frac{3-\epsilon _{0}}{1-\epsilon _{0}}\right)
^{2}+6\left( \frac{d\ln \epsilon }{dx}\right) _{0}-4R_{2}\right) ^{1/2},
\label{sol1}
\end{equation}
provided we treat $R_{2}$ as  constant -- this is essential to using a
Bessel function to represent the solution. However, $R_{2}$ is constant only
in the case when $\epsilon =$~constant, in which case $R_{2}$ itself is
precisely zero. The fact that $\epsilon $ and, hence, $R_{2}$ are $x$
-dependent for realistic models is what imposes a limit on the validity of
Bessel approximation and leads to the result that Bessel solution deviates
more and more from the exact solution as $|x|$ grows. For example, it tells
us that it is inappropriate to keep the terms smaller than $R_{2}$ in the
expression for index $\nu $. Consequently, keeping terms of order $\epsilon
_{0}^{2}$ in the expression for the index $\nu $ is invalid since these
terms are comparable to or smaller than the contributions of the $x$
-dependent terms in $R_{2}$ that have been neglected, unless $R_{2}$
vanishes. Eq.~(\ref{sol1}) should be re-expressed 
\begin{equation}
\nu \simeq \frac{3}{2}+\epsilon _{0}+\frac{1}{2}\left( \frac{d\ln \epsilon }{
dx}\right) _{0}+\delta _{2}\nu .  \label{sol2}
\end{equation}
where 
\begin{equation}
\delta _{2}\nu \sim \frac{5}{4}\epsilon _{0}^{2}-\frac{1}{3}R_{2}(x).
\label{acc}
\end{equation}
characterizes the ''accuracy'' with which the indexes in Bessel function
solution should be trusted.

Our analysis illustrates a key limitation to the horizon-crossing and Bessel
approximations: they cannot be improved further than first order by any
reasonable scheme. We have just argued that extending the approximation to
order $\epsilon_0^2$ requires that we also keep the $x$-dependent terms in $
R_2$; but then Eq.~ (\ref{sol1}) is no longer solvable in terms in Bessel
functions. Alternative schemes that we can imagine are as difficult as
solving the exact equations, and, hence, have no advantage.

The error in ignoring $\delta _{2}\nu $ translates into an error in the
Bessel approximation to the power spectrum that we wish to estimate. To
obtain the spectrum, we extend the Bessel function approximant valid about $
x=0$ to the long-wavelength ($x\gg 0$ or $\xi \rightarrow 0$) limit: 
\begin{equation}
H_{\nu }^{(1)}(\xi )=\frac{i\Gamma (\nu )}{\pi }\left( \frac{1}{2}\xi
\right) ^{-\nu }\left( 1-\frac{\xi ^{2}}{4(1-\nu )}+O(\xi ^{3})\right)
\label{as1}
\end{equation}
and the short-wavelength ($x\ll 0$ or $\xi \rightarrow \infty $) limit: 
\begin{equation}
H_{\nu }^{(1)}(\xi )=\sqrt{\frac{2}{\pi \xi }(P^{2}+Q^{2})}\exp \left(
i\left( \xi -\frac{\pi \nu }{2}-\frac{\pi }{4}+arctg\left( \frac{Q}{P}
\right) \right) \right)  \label{as2}
\end{equation}
where 
\begin{equation}
\begin{tabular}{l}
$P=1-\frac{(4\nu ^{2}-1)(4\nu ^{2}-9)}{2!(8\xi )^{2}}+O(\frac{1}{\xi ^{4}})$
\\ 
$Q=\frac{4\nu ^{2}-1}{8\xi }+O(\frac{1}{\xi ^{3}})$
\end{tabular}
\label{as3}
\end{equation}

Combining Eqs.~(\ref{def2}), (\ref{sol2}) and~(\ref{as1}) and  using $
H=H_{0}(1-\epsilon _{0}x+\ldots )$, the extrapolation of the Bessel solution
in the long-wavelength limit becomes 
\begin{eqnarray}
v &=&-iB_{k}\,z\,\,\sqrt{\frac{2}{\pi }}\frac{H_{0}}{k(\epsilon _{0})^{1/2}}
\left( 1-\beta \epsilon _{0}+\frac{1-\beta }{2}\left( \frac{d\ln \epsilon }{
dx}\right) _{0}\right)  \label{appr2} \\
&&\times \left( 1-\frac{\exp (-2x)}{4(1-\nu _{0})(1-\epsilon _{0})^{2}}
+O(e^{-3x})\right) \left( 1+O(\delta _{2}\nu )\right)  \nonumber
\end{eqnarray}
where $\beta =\ln 2+\gamma -1=0.27$ and $\gamma $ is Euler's constant, and $
B_{k}$ is a constant to be determined by matching to the short-wavelength
solution (to be discussed below). The last term in Eq.~(\ref{appr2})
characterizes the deviation of the Bessel solution from the exact solution
for $v$ based on the exact equation,  Eq. (\ref{bess}). This solution to the
approximant (Bessel) equation is to be matched to the long-wavelength limit
of the exact equation ($v=C(k)z$) to obtain $C(k)$ in terms of $B_{k}$, $
\epsilon _{0}$ and $(d{\rm ln}\,\epsilon /dx)_{0}$. We see that the match is
best if $x$ can be chosen so that the last two correction factors in
parentheses in Eq.~(\ref{appr2}) are negligible compared to $\beta
\,\epsilon _{0}$ and $\frac{1-\beta }{2}\,(d\,{\rm ln}\,\epsilon /dx)_{0}$.
That requires that $x$ be neither too small nor too big where the two
solutions are matched. The first correction factor requires that the
match-point $x=x_{+}$ satisfy $x_{+}\ge max\,\left\{ \left| {\rm ln}
(\epsilon _{0};\,\left( \frac{d{\rm ln}\epsilon }{dx}\right) _{0}\right|
\right\} $, and the second factor requires that $x_{+}$ be sufficiently
small that 
\begin{equation}
\delta _{2}\nu (x_{+})<<\min (\epsilon _{0};\left( \frac{d\ln \epsilon }{dx}
\right) _{0}).  \label{limit1}
\end{equation}
(Recall that $\delta _{2}\nu $ includes $x$-dependent terms that increase in
magnitude with increasing $x$.) The result is that 
\begin{equation}
x_{+}\simeq {\rm max}\left\{ \left| {\rm ln}\left[ {\rm min}(\epsilon
_{0};\,\left( \frac{d{\rm ln}\epsilon }{dx}\right) _{0}\right] \right|
\right\}  \label{limit2}
\end{equation}
is optimal for obtaining the best match. The fact that the optimal
match-point, $x_{+}$, is constrained from above and below means that there
is a residual second-order error which cannot be improved upon by
this matching procedure.

The matching of the Bessel solution to the short-wavelength limit proceeds
similarly and determines $B_{k}$. Eq.~(\ref{as2}) implies that 
\begin{equation}  \label{vB}
v=B_{k}\sqrt{\frac{2}{\pi }}\left( 1+O(1)\left( \frac{d\ln (1-\epsilon _{0})
}{dx}\right) _{0}x+\frac{1}{2}\exp (2x)+O(e^{4x})\right) \exp (i(k\tau ).
\label{sw}
\end{equation}
where the second terms inside the brackets comes as a result of uncertainty
in Bessel function index, $\delta_2 \nu$. 
This solution can be matched to the short-wavelength limit, $v\rightarrow 
\sqrt{1/2k}\,{\rm exp}(ik\tau )$ to first-order accuracy in $\epsilon _{0}$
and $(d{\rm ln}\,\epsilon /dx)_{0} $ at match-point $x_{-}\simeq \frac{1}{2}
\ln \left| \left( \frac{d\ln (1-\epsilon _{0})}{dx}\right) _{0}\right| ,$
roughly two or so e-folds before the end of inflation for typical models.
(Note that Eq.~(\ref{vB}) contains two second-order correction terms, one of
which increases with $x$ and the other which decreases with $x$.
Consequently, just as with $x_{+}$, we find that the point $x_{-}$, the
short-wavelength match-point which gives the best accuracy, is constrained
from above and below.) The higher-order corrections are negligible provided 
\begin{equation}
\delta _{2}B(x_{-})=O(1)\left( \frac{d\ln (1-\epsilon _{0})}{dx}\right)
_{0}\ln \left| \left( \frac{d\ln (1-\epsilon _{0})}{dx}\right) _{0}\right|
\ll 1.  \label{limit3}
\end{equation}
In this limit, $B_{k}=\sqrt{\pi /4k}(1+{\cal O}(1)\delta _{2}B(x_{-})$.
Substituting this expression for $B_{k}$ into the long-wavelength expression
Eq.~(\ref{appr2}) and obtaining a 
complete first-order expression for $C(k)$, 
one can then determine the power spectrum: 
\begin{equation}
P_{\zeta }(k)=\frac{k^{3}}{2\pi ^{2}}\left| \frac{v_{k}}{z}\right| ^{2}=
\left(\frac{H_{0}^{4}}{\dot{\varphi}}\right)_{0} \, 
\frac{1}{4 \pi^2} \left[ 1-2\beta \epsilon +(1-\beta
)\left( \frac{d\ln \epsilon }{dx}\right) +O(1)\delta _{2}B(x_{-})+O(1)\delta
_{2}\nu (x_{+})\right] _{0}  \label{power}
\end{equation}
where the last two terms in square
brackets characterize the uncertainty and should
be smaller than the previous, first-order terms. Comparing to the
original estimate, Eq.~(\ref{ex1}), we find that the ${\cal O}(1)$ factor
has been replaced by a function which depends on the equation-of-state.
 The spectral index of the
scalar power spectrum is then 
\begin{equation}
n_{s}\equiv 1+\frac{d\,{\rm ln}\,P_{\zeta }}{d\,{\rm ln}\,k}=
\left(1-2\epsilon 
-\frac{d\,{\rm ln}\,\epsilon }{dx}
-2 \epsilon^2 
-2\beta \epsilon \frac{d\,{\rm ln}
\,\epsilon }{dx}+(1-\beta )\frac{d^{2}\,{\rm ln}\,\epsilon }{dx^{2}}
\right)_{0}+\ldots  \label{index}
\end{equation}
and 
\begin{eqnarray}
\frac{dn_{s}}{dx} &=&(n_{s}-1)\frac{d(n_{s}-1)}{dx} \\
&=&\left( -2\epsilon \frac{d\,{\rm ln}\,\epsilon }{dx}
-\frac{d^{2}\,{\rm ln}
\,\epsilon }{dx^{2}}
-4\epsilon^2 \frac{d\,{\rm ln}\,\epsilon }{dx}
-2\beta \epsilon \left( \frac{d\,{\rm ln}\,\epsilon }{dx}
\right) ^{2}
-2\beta \epsilon \frac{d^{2}\,{\rm ln}\,\epsilon }{dx^{2}}
+(1-\beta )\frac{d^{3}\,{\rm ln}\,\epsilon }{dx^{3}}\right) _{0}+\ldots
\end{eqnarray}
where $\ldots $ are uncertainty due to $\delta _{2}\nu $ and $\delta _{2}B$.
These expressions agree with previous results\cite{Lidsey}$^,$\cite{LidTur}  except 
that they are expressed in terms of the equation-of-state and its derivatives.

Eqs.~(\ref{limit1}) through~(\ref{power}) summarizes the basic result: the
Bessel approximation for $P_{\zeta}(k)$ is only good to first order in $
\epsilon_0$ and $(d\, {\rm ln}\, \epsilon_0/dx)_0$ {\it at best}, and then
only if $\delta_2 B(x_-)$ and $\delta_2 \nu(x_+)$ are negligible compared to
the first order contributions. In previous discussions, it was pointed out
that the approximation was only valid if $\epsilon$ and $d \, {\rm ln} \,
\epsilon/dx$ (or equivalents) are nearly constant over some range of e-folds
around horizon-crossing,\cite{Lidsey}$^,$\cite{LidTur} but, otherwise, the conditions for the
approximation to be valid were not specified. Here, we see that the relevant
range of e-folds is between $x_-$ and $x_+$ e-folds about $k=aH$, typically
the five or so e-foldings surrounding horizon-crossing, $k=aH$. We also see
that, as $x\rightarrow x_{\pm}$, the Bessel solution approaches the exact
solution to within accuracy $\delta_2 \nu$ and $\delta_2 B$; for $x$ beyond
this range, the Bessel approximant diverges from the true short- and
long-wavelength solutions. Consequently, the Bessel approximant can achieve
first order accuracy, but no better. Recall that we must also restrict
ourselves to $d^2 \, {\rm ln} \, \epsilon/dx^2 \le {\cal O} (1)$ (see
discussion under Eq.~(\ref{first}).

In special cases, a satisfactory numerical result can be obtained even
though some constraints are not satisfied: A prominent example
is natural inflation and other potentials of the form $V\approx V_0
- a\varphi^2 + \ldots$.  In  Eq.~(\ref{limit2}), two independent
constraints are implied by 
 the parenthetical $\left (\epsilon _0;\left( \frac{d\ln \epsilon }{dx}
\right) _0\right)$.
The first constraint ensures that the
Bessel approximation gives the correct result to leading order in $
\epsilon_0 $; the second ensures that the correct result to leading order in 
$\left( \frac{d\ln \epsilon }{dx}\right) _0$. 
For some natural inflation models, though, the first constraint
is strongly violated so that the ${\cal O}(\epsilon_0)$ terms in the Bessel
approximation cannot be ``trusted." However, not only is the second
constraint satisfied, but the terms in the Bessel approximation proportional
to $(d\, {\rm ln}\, \epsilon/dx)_0$ are so much larger than the ${\cal O}
(\epsilon_0)$ terms that the violation of the first constraint is
numerically insignificant. The success of the Bessel approximation is
accidental in this sense (and may have deceived some into thinking that the
Bessel approximation has a much wider domain of validity than it actually
does).

Our elaborate analysis can be reduced  to a simple statement about
domain of validity of the horizon-crossing/Bessel
approximation: 

\noindent
(1) If $\epsilon=$
~constant, any value in the inflationary range between 0 and 1, the Bessel
solution is exact. However, a model with $\epsilon=$~constant is not
physically realistic since inflation never terminates. 

\noindent
(2)
If $\epsilon\ne$~constant, the Bessel approximation is only accurate if
 $\epsilon_0$ and $(d\, {\rm ln}\, \epsilon/dx)_0$
are small enough  that the first order
contributions in $\epsilon_0$ and $(d\, {\rm ln}\, \epsilon/dx)_0$ are much
larger than the higher order contributions, $\delta_2 \nu$ and $\delta_2 B$.
Suppose we demand that the higher order terms on left-hand-side of Eq.~(\ref
{limit2}) be less than a factor $\delta \ll 1 $ times the first order terms
on the right-hand-side of Eq.~(\ref{limit2}). (For example, $\delta$ might be
determined by the resolution of an experiment and we may wish to know if the
Bessel approximation provides the needed accuracy.) Assuming no accidental
cancellations,  Eqs.~(\ref{limit1}) through (\ref
{limit3}) reduce to: 
\begin{equation}  \label{reduce}
\begin{array}  {rcl}
\epsilon_0  & \le  &\frac{\delta}{x_+} \\
\left(\frac{d \, {\rm ln}\, \epsilon}{dx} \right)_0 & \le & \frac{\delta}{x_+} \\
\left(\frac{d^2 \, {\rm ln}\, \epsilon}{dx^2} \right)_0 &\le & \frac{2
\epsilon_0 \delta}{x_+} \le \frac{\delta^2}{x_+^2}; \, \, \ldots
\end{array}
\end{equation}
where $x_+ \equiv max \, \left\{ \left| {\rm ln} (\epsilon_0; \, \left(\frac{d {\rm 
n} \epsilon}{dx}\right)_0 \right| \right \} > 1$ and $\ldots$ refers to
analogous constraints on higher order derivatives. Even for modest accuracy, 
$\delta = 20\%$ and $x_+ \sim 2$, the ratio $\delta/x_+$ is $\sim
0.1 $, enough to highly restrict the range of $\epsilon$ and its derivatives.

Comparing 
the constraints above to 
 Eq.~(\ref{time}), one sees that the 
horizon-crossing/Bessel approximation
 applies to a wider range of models than the time-delay formalism.
Nevertheless, the range is narrow compared to full spectrum of inflationary
models.
One class of models in which the horizon-crossing/Bessel approximation is 
valid,
where $(d \, {\rm ln}\, \epsilon/dx)_0 \ll \epsilon_0 \le \delta/x_+ \ll 1$,
includes the simplest models of new inflation\cite{Linde,AS,ST}, chaotic
inflation\cite{Linde2} with $\phi^n$ potentials and $n>>2$, and extended
inflation\cite{LS,LS2}, which are realistic models incorporating inflation.
For these models, one obtains the CMB anisotropy prediction: $n_s$ and $r$
obey the relation\cite{Davis}: 
$r \simeq 21 (1+\gamma) \simeq 7 (1-n_s)$,
where $r\simeq \epsilon/14$ is the ratio of the tensor mode to the scalar
mode in terms of the contribution to the CMB dipole moment. The  second class,
where $\epsilon_0 \ll (d \, {\rm ln}\, \epsilon/dx)_0 \ll \delta/x_+ \ll 1$,
includes a range of natural inflation models,\cite{FFO} chaotic inflation
models with $\phi^n$ potentials and small $n$, and some two-field inflation
models\cite{AF} in which the inflaton field rolls near an extremum of the
potential during inflation. 

Some may have assumed that the good agreement
between the Bessel approximation and the exact methods for these two cases
 meant that the Bessel approximation 
could be used for a broader range of
models. In fact, our results show that these are essentially the only models 
for which the approximation can be trusted.

\noindent
{\bf Impact on   Microwave Background Anisotropy Prediction: An Illustration}
The error in using one of the approximate procedures instead of the more
cumbersome mode-by-mode integration propagates to predictions of the 
cosmic microwave background anisotropy and large-scale structure.
As a dramatic illustration,
 Figure~1 shows a comparison of the predicted
CMB anisotropy power spectrum  using the time-delay formalism
or naive horizon-crossing
approximation (based on Eq.~(\ref{ex1})), the Bessel approximation (based on
Eq.~(\ref{power})), and the exact computation for a sample inflaton
potential, $V(\phi)=\Lambda^4 
(1-\frac{2}{\pi}\, tan^{-1}(5 \phi/m_p)$,
 in which the equation-of-state changes rapidly enough near $\phi=0$ 
that Eq.~(\ref{reduce}) is not satisfied. 
(For this toy model, 
we have taken $\phi\approx -0.3$  to correspond to 60 e-folds before
the end of inflation.)
 The
discrepancy in the CMB predictions is large compared to the anticipated
experimental resolution of future space-based anisotropy experiments. 
Less dramatic effects occur in more typical models with slowly varying
equation-of-state;  an analysis for a wide spectrum of models  be presented
in a future paper.\cite{Fut1}  

\noindent
{\bf Summary:}
Our conclusions are summarized in     
Eqs.~(\ref{time}) and~(\ref{reduce}) as  constraints
on the equation-of-state, $\epsilon$.  The basic
result is that the time-delay and horizon-crossing methods are 
reliable approximations only if $\epsilon$ and its time-variation
are rather small.   These constraints can be  re-formulated in 
terms of rules-of-thumb for an inflaton potential, $V(\varphi)$:
 Assuming  higher-order corrections to our
approximation should be  $\delta < 0.20\%$  and $x_+ \approx 2$, 
then, if $V(\varphi)$ satisfies any of the following conditions
(recall that $4 \pi G =1$ and $x_+ > 2$):
\begin{eqnarray}
\left(\frac{V'}{V}\right)^2
        &\ge&  4\frac{\delta}{x_+} \approx 0.4 \label{V'constraint}\\
\frac{V''}{V}
        &\ge& \frac{\delta}{x_+} \approx 0.1\label{V''constraint} \\
\frac{V'V'''}{V^2} &\ge& \frac{\delta^2}{x_+^2} \approx 0.01
        \label{V'''constraint}
\end{eqnarray}
during the last 60 e-folds of inflation, 
the horizon-crossing/Bessel approximation  is not reliable and mode-by-mode 
integration is required.    For the time-delay formalism, the constraints
are roughly 60 times more stringent.

An important consequence  is that attempts at precise fitting of
CMB anisotropy data\cite{Kosowsky}$^{-}$\cite{Grivell} and large-scale
structure measurements and attempts to ``reconstruct" the inflaton potential
from CMB, as described in a recent review,\cite{Lidsey} is not as straightforward as one hoped. If the
horizon-crossing and reconstruction approaches were generally valid, then
inflationary predictions could be parameterized with only a few variables (
{\it e.g.}, $\epsilon_0$ and $(d \, {\rm ln} \, \epsilon/dx)_0$ evaluated
for the mode crossing the horizon in the present epoch). Simultaneous
fitting of these parameters along with other cosmic parameters, (such as the
Hubble constant, the cosmological constant, the baryon density, etc.) would
provide tight constraints on all. Indeed, this approach has been assumed in
most prior discussions of fitting data. To be sure, cases where the
equation-of-state is nearly constant and the horizon-crossing approximation
is valid appear to be the simplest forms of inflaton potential based on our
current understanding. So, one can decide {\it a priori} to assume this
subclass of inflationary potentials; in this case, there is no point to
general reconstruction methods since the potential forms are set by the {\it 
a priori} assumption. Alternatively, one may make no {\it a priori}
assumptions, in which case reconstruction methods are not useful since they
are not valid for general potentials. If we broaden the spectrum of possible
potentials, the fitting of cosmic parameters must be learned by comparing
data to some systematic search through exact results obtained by
mode-by-mode integration. How best to perform the search and how this
affects the empirical resolution of cosmic parameters from CMB measurements
is a subject of current investigation.\cite{Fut1}

This research was supported 
by the Department of Energy at Penn, DE-FG02-95ER40893 (LW and PJS), and
 by the Tomalla Foundation  (VM).

\newpage
\begin{figure}[h]
\epsfxsize=6 in \epsfbox{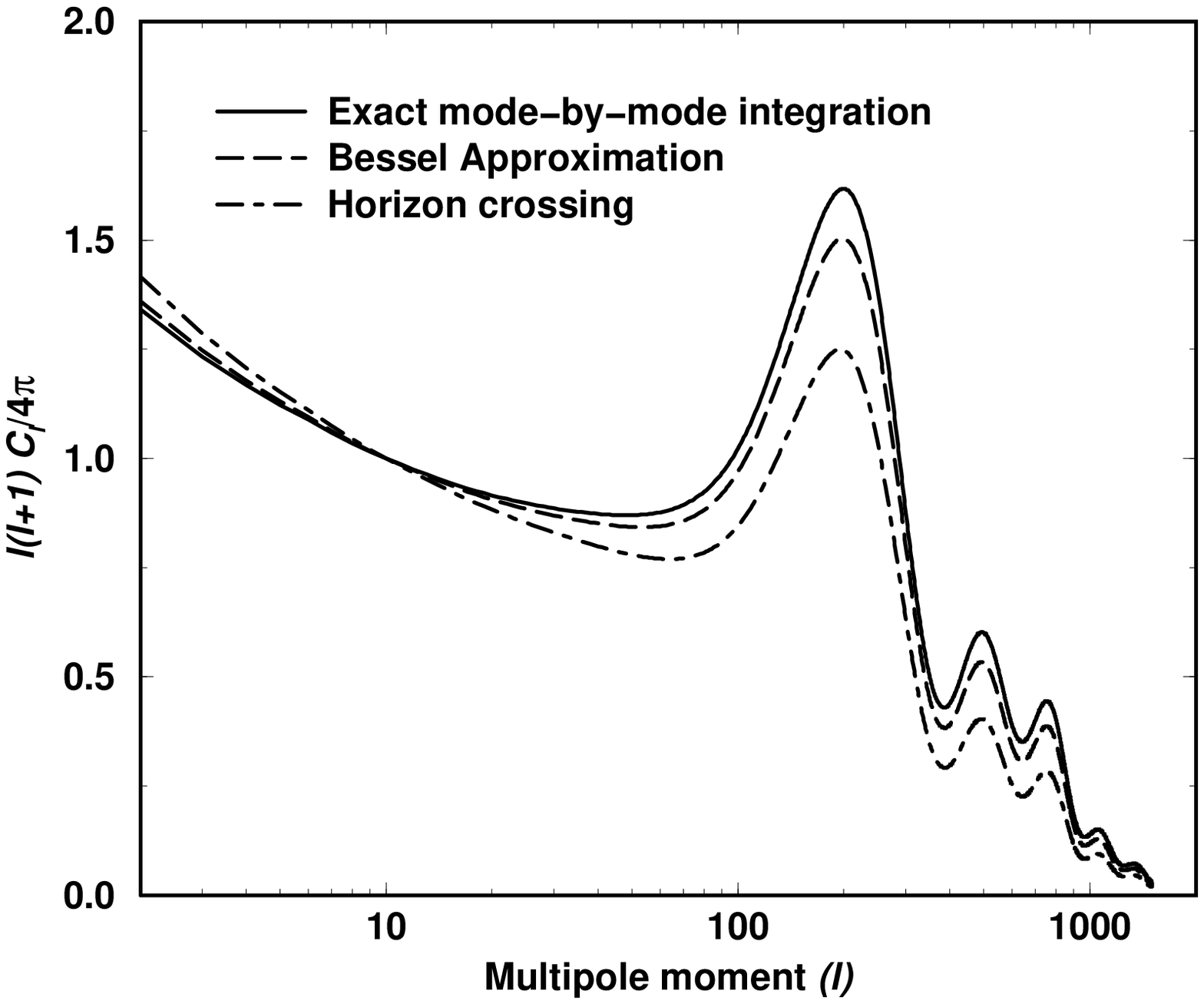}
\caption{ A comparison of the horizon-crossing and Bessel
approximations to  exact mode-by-mode integration for an inflaton
potential in which the
equation-of-state ($\epsilon$) is varying rapidly. 
The power spectrum has been
computed and converted into a prediction of the CMB temperature anisotropy
spectrum on large angular scales. }
\end{figure}

\end{document}